\documentclass[11pt,a4paper]{article}
\usepackage{amsmath,amsfonts,amssymb,amscd,graphicx}
\catcode`\@=11
\@addtoreset{equation}{section}

\catcode`\@=12

\newtheorem{Proposition}{Proposition}[section]

\newtheorem{Remark}{Remark}[section]

\def\Per{\otimes}

\def\v{\wedge}
\def\vv{\v\cdots\v}
\def\interior{\,\hbox{\vrule depth0pt height.6pt width4pt\vrule depth0pt height8pt}\;\,}

\def\tondo{\par\vskip6pt\hangindent.6cm\hangafter1\noindent$\;\;\bullet\;\;$}
\def\h#1{\hat{#1}}

\def\de#1/de#2{\frac{\partial {#1}}{\partial {#2}}}
\def\De#1/de#2{\textstyle{\text{\Large$\de{#1}/de{#2}$}}}
\def\sD#1/de#2/de#3{\textstyle{\text{\Large$\SD{#1}/de{#2}/de{#3}$}}}
\def\DE_#1{\partial_{\@#1}}

\def\@{\hskip.65pt}
\def\?{\hskip.3pt}
\def\plus#1#2{\vrule height#1pt width0pt depth#2pt}
\def\And{,\@\ldots\hskip-.4pt,}

\def\det{\operatorname{det}}

\def\Re{\mathbb R}

\def\a{\alpha}

\def\eps{\varepsilon}
\def\g{\gamma}
\def\l{\lambda}
\def\s{\sigma}

\def\w{\omega}

\def\z{\zeta}

\def\C{\mathcal C}
\def\D{\mathcal D}

\def\F{\mathcal F}
\def\G{\Gamma}

\def\L{\mathcal L}
\def\R{\mathcal R}
\def\T{\mathfrak T\?}

\def\W{\Omega}
\def\V#1{\mathcal V_{#1}}

\def\coord#1#2{#1^1\!\And #1^{#2}}
\def\trasf(#1,#2#3){#1=#1\/(\coord{#2}#3)\qquad i=1\@\and #3}

\def\GG#1(#2#3,#4){\ifx#1\h\h\G_{#2#3}{}^{#4}\else\G_{#2#3}{}^{#4}\fi}
\def\gg#1(#2#3,#4){\ifx#1\h\h\g\/_{#2#3}{}^{#4}\else\g\/_{#2#3}{}^{#4}\fi}
\def\ww#1(#2#3,#4){\if#1'\overline\w\/_{#2#3}{}^{#4}\else\w\/_{#2#3}{}^{#4}\fi}
\def\Chr(#1#2#3){\bigg\{\hbox to2.5ex{\hss$\begin{array}{c}
#1\\\hbox to0ex{\hss$#2\?#3$\hss}\end{array}$\hss}\bigg\}}

\begin{document}
\vskip-2cm

\title{Variational techniques in General Relativity\/: \\
       a metric affine approach to Kaluza's theory}

\author{Enrico Massa \\
        Dipartimento di Matematica, Universit\`a di Genova \\
        Via Dodecaneso 35. 16146 Genova (Italy) \\
                E-mail: massa@dima.unige.it
\and
        Stefano Vignolo\\
        DIPTEM Sez.~Metodi e Modelli Matematici, Universit\`a di Genova \\
                Piazzale Kennedy, Pad.~D. 16129 Genova (Italy)\\
                E-mail: vignolo@diptem.unige.it
}

\date{}
\maketitle

\begin{abstract}
A new variational principle for General Relativity, based on an action functional
$I\/(\Phi,\nabla)\/$ involving both the metric $\Phi\/$ and the connection $\nabla\/$ as
independent, \emph{unconstrained\/} degrees of freedom is presented. The extremals of $I\/$
are seen to be pairs $\/(\Phi,\nabla)\/$ in which $\Phi\/$ is a Ricci flat metric, and
$\nabla\/$ is the associated Riemannian connection. An application to Kaluza's theory of
interacting gravitational and electromagnetic fields is discussed.
\end{abstract}

\section{Introduction}
Several variational formulations of General Relativity, ranging from the purely metric
approach of Hilbert and Einstein \cite{Hilbert,Einstein} to Palatini's metric--affine
formulation \cite{Palatini,ADM,GRAV}, to the more recent purely affine \cite{Kijowski,
FK1,FK2} purely frame \cite{VCB1} and frame--affine theories \cite{FF,Rovelli,VC} have been
so far proposed in the literature

In particular, in the metric--affine formulation, the dynamical fields are pairs
$(\Phi,\nabla)\/$ consisting of a pseudo--riemannian metric $\Phi\/$ and of a torsionless
linear connection $\nabla\/$ on the space--time manifold $\V4\/$. The corresponding
variational principle relies on the action functional
\begin{equation*}
I\/(\Phi,\nabla)=\int g^{ij}R_{ij}\,\sqrt{|g|}\;d\/x^1\vv d\/x^4
\end{equation*}
where $g^{ij}\/$ are the contravariant components of the metric $\Phi\/$, and
$R_{ij}=R^h{}_{ihj}\/$ is the contracted curvature tensor associated with the connection
$\nabla$. The stationarity requirement for the functional $I\/$ singles out extremal pairs
$(\Phi,\nabla)\/$ in which $\Phi\/$ is a Ricci flat metric, and $\nabla\/$ is the associated
Riemannian connection.

In Palatini's approach, the absence of torsion, imposed as an a priori constraint, plays a
crucial role in the deduction of the field equations (for a generalization of this viewpoint
see e.g.~\cite{Massa}\@).

In this paper we propose an enhanced metric--affine principle, removing any restriction on
the choice of the connection. In the resulting scheme both the absence of torsion and the
condition $\@\nabla\Phi=0\/$ are part of the Euler--Lagrange equations associated with the
action functional.
The traditional Palatini--Hilbert and Einstein--Hilbert results are then recovered as
special cases of the more general procedure.

As an application of the new geometrical setup, in \S\;3 we discuss a variational approach
to Kaluza's theory of interacting electromagnetic and gravitational fields \cite{Kaluza,OW}.
The analysis relies on the introduction of a $5$--dimensional principal fiber bundle
$M\to\V4\@$ with structural group $(\Re,+)\/$, accounting for the gauge--theoretical
properties of the electromagnetic $4$--potential. Following Kaluza, we then merge the
gravitational and electromagnetic degrees of freedom into a symmetric tensor $\@\h\Phi\@$ of
signature $\@(4,1)\@$, playing the role of a metric tensor over $\/M$. We finally show that
this metric, together with the associated Levi--Civita connection, are the extremals of a
constrained variational problem of the proposed kind.
\\
An advantage of the new formulation is that it involves only the physical (gravitational and
electromagnetic) fields, and does not require any additional geometric object, such as the
scalar field reported in \cite{OW}.

\section{The action principle}
\subsection{Mathematical preliminaries}
Let $M\/$ be an $n$--dimensional orientable manifold, $\R\/(M)\xrightarrow{\pi}M\/$ the
bundle of symmetric covariant tensors of rank $2\/$ and signature $(p,q)\/$ over $M\/$, and
$\C\/(M)\xrightarrow{\pi}M\/$ the bundle of linear connections over $M\/$.
\\
The existence of \emph{global\/} sections $\Phi:M\to \R\/(M)\/$ is explicitly assumed. Each
such section is called a \emph{pseudo--riemannian metric\/} on $M\/$.
\\
We refer $M\/$ to local coordinates $(U,x^1,\ldots,x^n)\/$ and adopt a (possibly non
holonomic) basis $\@\{\@\DE_i\,,\@\w^i,\,i=1\And n\,\}\/$ for the tensor algebra over $U\/$.
The latter induces fiber coordinates on $\R\/(M)\/$ and $\C\/(M)\@$, respectively denoted by
$\@x^i,\@y_{ij}\@$ and $\@x^i,\@\gg_(ij,k)\/$.

\noindent
The following results will be regarded as known:
\begin{itemize}
\item
$\C\/(M)\xrightarrow{\pi}M\@$ is an affine bundle, modelled on the bundle $\@\T^1_2\/(M)\@$
of tensors contravariant of degree $1$ and covariant of degree $2$. In particular,
$\C\/(M)\@$ always admits \emph{global\/} sections $\@\nabla:M\to\C\/(M)\@$. Each such
section, locally represented as $\@\gg(ij,k)=\GG(ij,k)\/(\coord xn)\@$, is called a
\emph{connection\/} over $M\/$.
\\
For any $\@X\in D^1\/(M)\@$, we denote by $\@\nabla\!_X\@$ the \emph{covariant derivative\/}
along $X\/$ induced by $\nabla$, namely the derivation of the tensor algebra $\D\/(M)\/$
depending $\@\F$--linearly on $X\/$ and commuting with contractions, uniquely determined by
the requirement
\begin{equation}\label{2.1}
\nabla\!_X\/(f)=X\/(f) \,,\qquad \nabla_{\!\DE_i}\,\DE_j = \GG(ij,k)\;\DE_k
\end{equation}

\item
$\C\/(M)\@$ carries an affine surjection $T\/$, known as the \emph{torsion map\/}, into the
subbundle of $\T^1_2\/(M)\@$ formed by the totality of tensors antisymmetric in the
covariant indices. In local coordinates, denoting by
$\,C\@^i{}_{jk}:=\big<\,[\@\DE_j\@,\@\DE_k\@]\,,\@\w^i\@\big>\,$ the holonomy tensor of the
basis $\{\@\DE_i\,,\@\w^i\@\}\/\plus0{3.5}$, we have the explicit representation
$\@T\/(\G)=T^i{}_{jk}\@\big(\DE_i\Per\w^j\Per\w^k\big)_{\pi\/(\G)}\@$, with
\begin{equation}\label{2.2}
T^i{}_{jk}\,=\,\gg(jk,i) - \gg_(kj,i) - C^i{}_{jk}
\end{equation}

\item
Assigning a pseudo--riemannian metric $\Phi :M\to \R\/(M)\/$ singles out a distinguished
section $\h\nabla:M\to\C\/(M)\/$, called the \emph{riemannian connection\/} of $\Phi\/$. The
latter determines a bijection of $\C\/(M)\/$ into the modelling space $\T^1_2\/(M)\@$
assigning to each $\G\in\C\/(M)\/$ the difference $N\/(\G):=\G-\h{\nabla}_{|\pi\/(\G)}\@$.
Denoting by $\GG\h(jk,i)\/(x^1\And x^n)\/$ the connection coefficients of $\h\nabla\/$ in
the basis $\@\{\@\DE_i\,,\@\w^i\@\}\/$, the image $\@N\/(\G)\@$ is locally represented as
$\@N^i{}_{jk}\,(\DE_i \otimes\w^j\otimes\w^k)_{\pi\/(\G)}\@$, with
\begin{equation}\label{2.3}
N^i{}_{jk}=\gg(jk,i) - \GG\h(jk,i)
\end{equation}
In terms of $N\/$, eq.~(\ref{2.2}) provides the identification
\begin{equation}\label{2.4}
T^i{}_{jk} = N^i{}_{jk} - N^i{}_{kj}
\end{equation}
\end{itemize}
The fibered product $\R\/(M)\times_M\C\/(M)\/$ is the natural environment for the development of
a field theory in which every global section $M\to\R\/(M)\times_M \C\/(M)\@$ corresponds to the
simultaneous assignment of a pseudo--riemannian structure $\Phi\/$ and of a connection
$\nabla\/$ over $M\/$. This is precisely the viewpoint we shall pursue. The field theory we
shall discuss relies on the action functional
\begin{equation}\label{2.5}
I\/(\phi,\nabla)\,:=\,\int_D\@g^{ij}\left(R_{ij}+ T_i\@T_j\right)\,\sqrt{|g|}\,\w^1\vv\w^n
\end{equation}
$R_{ij}:=R\@^p{}_{ipj}\@$ and $\@T_i :=T\@^p{}_{pi}\@$ respectively denoting the contracted
curvature tensor and the contracted torsion tensor of the connection $\nabla\/$. We shall
prove that the extremals of the functional \eqref{2.5} are pairs $(\Phi,\nabla)\/$ such that
\begin{itemize}
\item
$\nabla\/$ is the riemannian connection of $\Phi\@$;
\item
the metric $\Phi\/$ is ``Ricci flat'', i.e.~it obeys Einstein's equation \emph{in vacuo\/}
\begin{equation*}
R_{ij} =0
\end{equation*}
\end{itemize}

\subsection{The field equations}
To fulfill our program, we refer $\R\/(M)\times_M\C\/(M)\/$ to coordinates
$x^i,y_{ij},\gg(ij,k)\/$. Every section $\@(\Phi,\nabla):M\to\R\/(M)\times_M\C\/(M)\/$ is then
described locally as
\begin{equation}\label{2.6}
y_{ij}\,=\,g_{ij}\/(x^1\And x^n)\;,\qquad\gg_(ij,k)=\GG(ij,k)\/(x^1\And x^n)
\end{equation}
We denote by $\w^i{}_j:=\GG(kj,i)\,\w^k\/$ the connection $1$--forms of $\@\nabla\@$ in the
basis $\{\@\DE_i\,,\@\w^i\@\}\/\plus13$, and by $\theta\@^i:=\frac12
T^i{}_{jk}\,\w^j\v\w^k\@$ and $\rho\?^i{}_j:=\frac12R^i{}_{jkl}\,\w^k\v\w^l\/\plus13$
respectively the torsion $2$--forms and the curvature $2$--forms of $\@\nabla\/$. The
relationships between the various objects are summarized into Cartan's structural equations
\begin{subequations}\label{2.7}
\begin{equation}
\theta\@^i\,=\,d\@\w^i + \w^i{}_p\v\w\@^p
\end{equation}
(pointwise equivalent to eq.~(\ref{2.2})) and
\begin{equation}
\rho\?^i{}_j\,=\,d\@\w^i{}_{j} + \w^i{}_{p}\v\w\@^p_{\;\,j}
\end{equation}
\end{subequations}
In terms of $\theta\@^i\/$ and $\rho\?^i{}_j\/$, the contracted torsion and curvature
tensors involved in eq.~(\ref{2.5}) are respectively expressed by the relations
\begin{equation}\label{2.8}
T_i\,=\,\big<\,\DE_p\v\DE_i\,\@|\,\@\theta\,^p\,\big>\;,\qquad
R_{ij}\,=\,\big<\,\DE_p\v\DE_j\,\@|\,\@\rho\,^p{}_i\,\big>
\end{equation}
We keep the notation $\h\nabla\/$ for the Riemannian connection of $\Phi\/$, and denote by a
hat all quantities pertaining to $\h\nabla\/$ (connection coefficients, connection
$1$--forms, etc.).

According to eq.~(\ref{2.3}), the relation between the connection $1$--forms of $\nabla\/$
and those of $\h\nabla\/$ is locally expressed as
\begin{equation}\label{2.9}
\w^i{}_{j}\,=\,\h\w^i{}_j + N^i{}_j
\end{equation}
with $N^i{}_j:= N^i{}_{kj}\,\w\?^k = (\@\GG(kj,i) -\GG\h(kj,i)\@)\,\w\?^k\/$.
\\
On account of eqs.~(\ref{2.7}a, b), this yields the identifications
\begin{subequations}\label{2.10}
\begin{align}
& T_i\,=\,\big<\,\DE_p\v\DE_i\,\@|\,\@N\@^p{}_q\,\w\?^q\,\big>\,=
\,\delta\@^{rq}_{pi}\,N^p{}_{rq}                                                \\[3pt]
& R_{ij}\,=\,\big<\,\DE_p\v\DE_j\,\@|\,\@\h\rho\,^p{}_i+ d\?N\@^p{}_i +
N\@^p{}_q\v\h\w\?^q{}_i + \h\w\@^p{}_q\v N\@^q{}_i + N\@^p{}_q\v N\@^q{}_i\,\big>
\end{align}
\end{subequations}
On the other hand, a straightforward computation provides the relation
\[
d\?N\@^p{}_i + N\@^p{}_q\v\h\w\?^q{}_i + \h\w\@^p{}_q\v N\@^q{}_i \,=\,
\h{\nabla}_{\!\@\DE_k}\,N\@^p{}_{ri}\;\w^k\v\w^r
\]
Collecting all results, we end up with the expression
\begin{multline}\label{2.11}
g^{ij}\left(R_{ij}\@+\@T_iT_j\right)\,=\,g^{ij}\/\left[\h{R}_{ij} + \delta\@^{kr}_{pj}\/
\left(\h{\nabla}_{\!\@\DE_k}\,N\@^p{}_{ri} + N\@^p{}_{kq}\,N\@^q{}_{ri}\@\right)\@
+\@T_i\,T_j
\right]\,=                                                                          \\
=\,g^{ij}\/\left(\h{R}_{ij} + \delta\@^{kr}_{pj}\@N\@^p{}_{kq}\,N\@^q{}_{ri} +
T_i\,T_j\right) + g^{ij}\/\left(\h{\nabla}_{\!\@\DE_p}\,N\@^p{}_{ji} -
\h{\nabla}_{\!\@\DE_j}\,N\@^p{}_{pi}\right)
\end{multline}
This shows that, up to a divergence, the action functional (\ref{2.5}) may be written in the
equivalent form
\begin{equation}\label{2.12}
I\/(\Phi,\nabla)\,=\,\int_D\,g^{ij}\/\left(\h{R}_{ij} + \delta\@^{kr}_{pj}\@
N\@^p{}_{kq}\,N\@^q{}_{ri} + T_i\,T_j\right)\sqrt{|g|}\,\w^1\v\dots\v\w^n
\end{equation}
with $T_i\/$ given by eq.~(\ref{2.10}a) and with $\h{R}_{ij}=\h R^{\@\@p}{}_{ipj}\@$
representing the \emph{Ricci tensor\/} associated with the metric $\Phi\/$.

Both expressions (\ref{2.5}), (\ref{2.12}) have their own advantages: eq.~(\ref{2.5})
depends algebraically on $\Phi\/$, thereby allowing a simple description of the variation of
the functional $I\/$ under arbitrary deformations of the metric. On the contrary,
eq.~(\ref{2.12}) depends algebraically on $\@\nabla\/$, thus yielding an equally simple
expression for $\@\delta I\@$ under arbitrary deformations of the connection. Let us work
out both aspects in detail.
\\[1pt]
1) \,On account of the relation
\begin{equation}\label{2.13}
\de\@\sqrt{|g|}/de{g^{ab}}\,=\,\de/de{g^{ab}}\,\frac1{\@\sqrt{|g|^{-1}}}\,=\,-\frac12\,
\big|\@g\@\big|^{\?\frac32}\,\@\de g^{-1}/de{g^{ab}}\,=\,-\frac12\,\sqrt{|g|}\;g_{ab}
\end{equation}
the variation of $I\/$ under arbitrary deformations $\@\delta g^{ab}\@$ takes the form
\begin{equation}\label{2.14}
\delta I=\int_D\/\left[R_{ab}+T_a\@T_b-\frac12\@\left(R+T_p\@T\@^p\@\right)g_{ab}\right]\/
\delta g^{ab}\,\sqrt{|g|}\;\w^1\v\dots\v\w^n
\end{equation}
with $\@R=g^{ab}\@R_{ab}\@$ and $\@T\?^p=g\@^{pq}\@T_q\@$.
\\
In the case of \emph{unconstrained\/} deformations, the requirement $\delta I=0\/$ is
therefore expressed by the condition
\begin{equation}\label{2.15}
R_{ab}\,+\,T_a\@T_b\,-\,\frac12\@\?\big(R + T_p\@T\@^p\@\big)\,g_{ab}\,=\,0
\end{equation}
In dimension $n>2\@$ the latter reduces to
\begin{equation}\label{2.16}
R_{ab}\,+\,T_a\@T_b\,=\, 0
\end{equation}
If the class admissible metrics is restricted to a subfamily $\@\Phi\/(\xi^1\And\xi^r)\@$
controlled by a smaller number of fields, eq.~(\ref{2.14}) is still valid, but
eq.~(\ref{2.15}) is replaced by the system
\begin{equation}\label{2.17}
\left[R_{ab}\,+\,T_a\@T_b - \frac12\@\?\big(R\,+\,T_p\@T\@^p\@\big)\,g_{ab}\right]
\de{g^{ab}}/de{\xi^j}\,=\,0
\end{equation}
An example of this situation will be illustrated in Section 3.

\smallskip\noindent
2) \,In order to evaluate the variation $\@\delta\@I\/$ under arbitrary deformations of the
connection we resort to the representation (\ref{2.12})\@. From the latter, making use of
the identifications $\@\delta\@\GG(bc,a)=\delta N^a{}_{bc}\,$,
$\@\delta\@\?T_i=\delta\@^{bc}_{ai}\,\@\delta N^a{}_{bc}\/$ we get the expression
\begin{equation}\label{2.18}
\delta I=\int_D\left[\delta\@^{kr}_{pj}\/\Big(\delta\?^p_a\,\delta^b_k\,N^c{}_r{}^j +
\delta\@^b_r\,g^{jc}\,N^p{}_{ka}\Big)+2\@\delta\@^{bc}_{ai}\,\@T\@^i\right]\!\@
\delta N^a{}_{bc}\,\sqrt{|g|}\;\w^1\v\dots\v\,\w^n
\end{equation}
In the case of unconstrained deformations $\delta N^a{}_{bc}\,$, the requirement $\delta I=
0\@$ is therefore expressed by the condition
\begin{equation}\label{2.19}
0\,=\,\delta\@^{br}_{aj}\,N^c{}_r{}^j\@+\,\delta\@^{kb}_{pj}\,N^p{}_{ka}\,g^{jc}\@+\,
2\,\delta\@^{bc}_{ai}\;T\@^i
\end{equation}
From the latter, contracting $a\/$ with $c\/$, we derive the relation
\begin{equation}\label{2.20}
0\,=\,-\@2\,(n-1)\,T\@^b \qquad \Longrightarrow \qquad T\@^b\@=\,0
\end{equation}
In view of this, eq.~(\ref{2.19}) reduces to
\begin{equation}\label{2.21}
0\,=\,\delta\@^b_a\,N^c{}_r{}^r\@-\,N^c{}_a{}^b\@+\,N^r{}_{ra}\,g^{bc}\@-\,N^{bc}{}_a
\end{equation}
Setting $\,X^c:=N^c{}_r{}^r,\,\@Y_a:=N^r{}_{ra}\@$ and lowering all indices,
eq.~(\ref{2.21}) takes the form
\begin{equation*}
N_{cab}\,+\,N_{bca}\,=\,g_{ab}\,X_c\,+\,g_{bc}\,Y_a
\end{equation*}

\noindent
The latter is easily solved for $\@N_{cab}\@$, yielding the expression
\begin{equation*}
2\@N_{cab}\,=\,g_{ab}\,(\?X_c\@-\@Y_c\?)\,+\,g_{ac}\,(\?Y_b\@-\@X_b\?)\,+
\,g_{bc}\,(\?Y_a\@+\@X_a\?)
\end{equation*}

\noindent
From this, recalling eqs.~(\ref{2.10}\@a), (\ref{2.20}) as well as the definition of
$\@Y_a\@$, we get the relations
\begin{equation*}
\begin{alignedat}{2}
2\@Y_b\,&=\,2\@g^{ac}\,N_{cab}\,=\,X_b\@+\@n\@(Y_b - X_b)\@+\@X_b \qquad &&\Rightarrow
\qquad (n-2)\@(Y_b - X_b)\,=\,0                                                      \\[4pt]
0\,&=\,T_a\,=\,g^{bc}\@(N_{cab}-N_{cba})\,=\,(n+1)\@Y_a\qquad&&\Rightarrow \qquad Y_a\,=\,0
\end{alignedat}
\end{equation*}

\noindent
Collecting all results we conclude that, for $n>2\@$, the requirement $\delta I=0\/$ is
mathematically equivalent to $\@N^i{}_{jk}=0\@$, i.e.~to the identification
$\@\nabla=\h\nabla\@$.
\\
This fact, together with eq.~(\ref{2.16}), provides a full proof of the result stated in
\S~2.1.
\begin{Remark}\label{Rem2.1}\rm
Since being an extremal with respect to a class $\@\mathfrak E\@$ of deformations
automatically implies being an extremal with respect to any subclass $\@\mathfrak E'\subset
\mathfrak E\@$, the consequences of the variational principle based on the functional
(\ref{2.5}) hold unchanged if part of the conditions arising from the requirement
$\@\nabla=\h\nabla\@$ are imposed as a priori constraints. Thus, for example, if the choice
of $\nabla\/$ is restricted to the class of torsionless connections, the previous analysis
provides a proof of the Palatini--Hilbert action principle.

\noindent
More radically, if one gives up the affine degrees of freedom and considers a purely metric
setup, with the ansatz $\@\nabla=\h\nabla\@$ imposed at the outset, the action principle
(\ref{2.5}) is easily recognized to yield back the Einstein--Hilbert one.
\end{Remark}

\section{Affine scalars and the Einstein--Maxwell theory}
As an illustration of the results developed so far we discuss an application of the
functional (\ref{2.5}) to the study of the Einstein--Maxwell equations. The argument
provides a geometric approach to Kaluza's theory of interacting gravitational and
electromagnetic fields, free of any spurious, non--physical field (see e.g.~\cite{OW} and
references therein).

Let $\@\V4\@$ denote a $4$--dimensional orientable space--time manifold, admitting a
pseudo--riemannian structure of signature $(3,1)\/$. Also, let $\@M\xrightarrow{\pi}\V4\@$
denote a principal fiber bundle over $\@\V4\@$ with structural group $(\Re,+)\@$, henceforth
referred to as the bundle of \emph{affine scalars\/}.

The bundle $M\to\V4\@$ is globally trivial. Assigning a trivialization $\@u:M\to\Re\@$
allows to lift every coordinate system $\@x^1\And x^4\/$ in $\@\V4\@$ to a corresponding
fibered coordinate system $\@u,x^1\And x^4\/$ in $M\@$. The group of fibered coordinate
transformations has then the form
\begin{equation}\label{3.1}
\bar u\,=\,u + f\/(\coord x4)\;,\qquad \bar x^i\,=\,\bar x^i\/(\coord x4)
\end{equation}
In fibered coordinates, the generator of the action of $\@(\Re,+)\@$, commonly referred to
as the \emph{fundamental vector field\/} of $M\/$, coincides with the field $\@\DE_u:=
\de/de u\@$.

The presence of $\@\DE_u\@$ singles out a distinguished sub--bundle
$\@\h\R\?(M)\xrightarrow{}M\@$ of the bundle of pseudo--riemannian structures of signature
$(4,1)\/$ over $\@M$, formed by the totality of metrics satisfying the condition
$\@\big(\?\DE_u\@,\@\DE_u\big)=1\@$.

Through an obvious composition of maps, $\@\h\R\?(M)\@$ may be viewed as a fiber bundle over
$\?\V4\@$. In the resulting context, assigning a section $\@\Upsilon:\V4\to
\h\R\?(M)\@$ is then equivalent to assigning a pair $\@(\psi,\h\Phi)\@$ where
\begin{itemize}
\item
$\@\psi:\V4\to M\@$ is a section, described locally as $\,u=\psi\/(\coord x4)\@$;
\item
$\@\h\Phi:M\to\h\R\/(M)\@$ is a pseudo--riemannian metric on $\?M\/$, uniquely characterized
by the requirements
\begin{subequations}\label{3.2}
\begin{equation}
\h\Phi\@_{|\@\psi\/(x)}\,=\,\Upsilon\/(x)\quad\forall\;x\in\V4\,,\hskip2.4cm
\L_{\@\DE_u}\h\Phi\,=\,0    \hskip.8cm
\end{equation}
locally summarized into the representation
\begin{equation}
\h\Phi\,=\,d\/u\Per d\/u\@+\@2\,\g_i\/(\coord x4)\,d\/u\odot d\/x^i\@+\@
\g_{ij}\/(\coord x4)\,d\/x^i\Per d\/x^j\hskip.4cm
\end{equation}
\end{subequations}
\end{itemize}
\noindent
Denoting by the $g:T\/(M)\to T^*\/(M)\/$ the process of ``lowering the indices'' induced by
the metric (\ref{3.2}\@b)\@, let us now consider the $1$--form
\begin{subequations}\label{3.3}
\begin{equation}
\s\,:=\,g\/\big(\DE_u\big)\,=\,d\/u\@+\@\g_i\,d\/x^i
\end{equation}
By direct computation we have then the relations $\big<\s\@,\@\DE_u\@\big>=1\@,\,
\L_{\@\DE_u}\s=0\@$, indicating that $\@\s\@$ defines a \emph{principal connection}
relative to the fibration $\@M\to\V4\@$. For each choice of the section $\@\psi\@$, the
knowledge of $\@\s\@$ is therefore equivalent to the knowledge of the pull--back
$\@\psi^{\@*}\/(\s)\in\D_1\/(\V4)\,$.

In a similar way, in view of eqs.~(\ref{3.2}\@b), (\ref{3.3}\@a)\@, the difference
$\@\h\Phi\@-\@\s\Per\s\@$ is easily recognized to coincide with the pull--back of a tensor
field over $\@\V4\@$, expressed in coordinates as
\begin{equation}
\Phi\,=\,\big(\@\g_{ij}\@-\@\g_i\,\g_j\@\big)\,d\/x^i\Per d\/x^j\,:=\,g_{ij}\,
d\/x^i\Per d\/x^j
\end{equation}
\end{subequations}
Collecting all results, we have therefore the representation
\begin{equation}\label{3.4}
\h\Phi\,=\,\s\Per\s\@+\@\pi^*\/(\Phi)
\end{equation}

The idea is now to interpret the tensor (\ref{3.3}\@b) as a pseudo--riemannian metric on
$\@\V4\@$, accounting for the gravitational effects, and to regard the connection $\@\s\@$
as a description of the electromagnetic field. In this way, up to a dimensional constant
$\?\kappa\@$ depending on the choice of the units, every section
$\@\Upsilon:\V4\to\h\R\/(M)\@$ is identified with a triple $\@(\psi,A,\Phi)\,$ where\@:
\tondo $\@\psi:\V4\to M\@$ is a section, accounting for the electromagnetic gauge;
\tondo $\@A:=\kappa^{-1}\,\psi^*\/(\s)\@$ is the ($\@\psi$--dependent)
electromagnetic $\@4$--potential in $\@\V4\@$;
\tondo $\@\Phi=g_{ij}\,d\/x^i\Per d\/x^j\@$ is the ($\@\psi$--independent) metric tensor of
$\@\V4\@$.

\bigskip
The algorithm is significantly simplified referring the tensor algebra $\@\D\/(M)\@$ to the
local non--holonomic basis $\{\@\DE_A\@,\@\w^A,\;A=0\And 4\@\}\,$ defined by the ansatz
\begin{equation}\label{3.5}
\w^0=\sigma\@,\quad\; \w^i=d\/x^i\,,\quad\; \DE_0 =\DE_u=\de/de u\,,
\quad\; \DE_i=\de/de{x^i}\,-\@\g_i\,\de/de u
\end{equation}
In view of eqs.~(\ref{3.4}), (\ref{3.5}), the covariant representation of the metric reads
\begin{subequations}\label{3.6}
\begin{equation}
\h\Phi\,:=\,g_{AB}\;\w^A\Per\w^B\,=\,\w^0\Per\w^0\@+\@g_{ij}\;\w^i\Per\w^j
\end{equation}
In a similar way, the contravariant representation of $\@\h\Phi\@$ takes the form
\begin{equation}
g^{AB}\,\DE_A\Per \DE_B = g^{ij}\,\DE_i\Per \DE_j + \DE_0\Per\DE_0
\end{equation}
\end{subequations}
with $g^{ij}g_{jk} = \delta^i_k\@$. Eqs.~(\ref{3.6}\@a, b) imply the identity

\begin{equation}\label{3.7}
1\,=\,g^{00}\,=\,\frac{\det g_{ij}}{\det g_{AB}}
\end{equation}
Setting $\@\det g_{AB}=\det g_{ij}:=g\@$, and denoting by
$\@\h\eps:=\sqrt{|g|}\;\w^0\vv\w^4\@$\vspace{1pt} and $\@\eps:=\sqrt{|g|}\;d\/x^1\vv
d\/x^4\@$ the Ricci tensors respectively associated with the metrics $\@\h\Phi\@$ and
$\@\Phi\@$, eqs.~(\ref{3.3}\@a), (\ref{3.5}), (\ref{3.7}\@a) yield the relations
\begin{subequations}\label{3.8}
\begin{align}
& \h\eps\,=\,\w^0\v\pi^*\/(\eps)\,=\,\sqrt{|g|}\;d\/u\v d\/x^1\vv d\/x^4        \\[3pt]
& \DE_0\interior\h\eps\,=\,\pi^*\/(\eps)\,=\,\sqrt{|g|}\;d\/x^1\vv d\/x^4
\end{align}
\end{subequations}

Given any section $\@\Upsilon=(\psi,\h\Phi):\V4\to\h\R\?(M)\@$, let $\@\h\nabla\@$ denote
the riemannian connection of $\@\h\Phi\@$. The 4+1 decomposition expressed by
eq.~(\ref{3.4}) is then reflected into an analogous representation of $\@\h\nabla\@$ in
terms of the Christoffel symbols of $\@\Phi\@$ and of the curvature $2$--form of $\@\s\@$.
Setting
\begin{subequations}\label{3.9}
\begin{align}
& \h\w\@^A{}_B:=\GG\h(CB,A)\,\w\@^C,\qquad\;\w^{\@*\,i}{}_j:=\Chr(ikj)\,\w^k,\qquad\;
\Omega:=d\?\w\@^0=\Omega\@_{ij}\,\w^i\v\w^j                                         \\
\intertext{a straightforward calculation yields the results}
& \h\w\@^i{}_j\,=\,\w^{\@*\,i}{}_j\,-\,\W\@^i{}_j\,\w\@^0,\quad\;
\h\w\@^i{}_0\,=\@-\,\W\@^i{}_j\,\w\@^j,\quad\;\h\w\@^0{}_i\,=\,\W\@_{ij}\,\w\@^j,\quad\;
\h\w\@^0{}_0\,=\,0
\end{align}
\end{subequations}

\smallskip
To complete our geometrical setup let us finally denote by $\@\h\R\/(M)\times_M\@\C\/(M)\,$ the
fibered product of $\@\h\R\/(M)\@$ with the bundle of linear connections over $\@M$, and by
$\@p_1:\h\R\/(M)\times_M\@\C\/(M)\to\h\R\/(M)\@$, $\@p_2:\h\R\/(M)\times_M\@\C\/(M)\to\C\/(M)\@$ the
associated natural projections. Once again, we regard $\@\h\R\/(M)\times_M\@\C\/(M)\@$ as a fiber
bundle over $\@\V4\@$. Assigning a section $\@\Xi:\V4\to\h\R\/(M)\times_M\@\C\/(M)\@$ is then
easily recognized to be mathematically equivalent to assigning a triple
$\@(\psi,\h\Phi,\nabla)\@$, where:
\begin{itemize}
\item
the pair $\@(\psi,\h\Phi)\@$ is defined exactly as above, with the section
$\@\Upsilon:\V4\to\h R\/(M)\@$ now identified with the product $\@p_1\cdot\Xi\,$;
\item
$\@\nabla:M\to\C\/(M)\@$ is a linear connection over $M\/$, invariant under the action of
the structural group $\/(\Re,+)\@$ and satisfying
$\@\nabla\@_{|\@z}=p_2\cdot\Xi\,(\pi\@(z)\?)\;\forall\;z\in\psi\/(\V4)\@$. As intuitively
clear this means that, in the non--holonomic basis (\ref{3.5})\@, $\@\nabla\@$ is described
by connection coefficients $\@\GG(AB,C)\@$ \emph{independent of the variable\/} $\@u\?$
\footnote{A review of the concept of Lie derivative of a connection is reported in
Appendix A.}.
\end{itemize}

After these preliminaries, let us now adapt the variational scheme of \S\;2 to context in
study. To this end, to every section $\@\Xi:\V4\to\h\R\/(M)\times_M\@\C\/(M)\@$, viewed as a
triple $\@(\psi,\h\Phi,\nabla)\@$ in the sense described above, we associate the action
functional
\begin{equation}\label{3.10}
I\/(\?\Xi\?)\,:=\,\int_{\psi\/(D)}\,g^{AB}\left(R_{AB}+T_A\,T_B\right)
\@\sqrt{|g|}\,d\/x^1\vv d\/x^4
\end{equation}
$D\@$ being any domain with compact closure in $\@\V4\@$, and $\@\sqrt{|g|}\;d\/x^1\vv
d\/x^4\@$ denoting the invariant $4$--form (\ref{3.8}\@b)\@.

Up to straightforward notational changes, evaluating of the right hand side of
eq.~(\ref{3.10}) involves the same type of algorithm already exploited in \S\,\@2\@. In
particular eq.~(\ref{2.11}) takes now the form
\begin{multline}\label{3.11}
g^{AB}\/\Big(R_{AB}+T_A\@T_B\Big)\,=\,g^{AB}\/\Big(\h{R}_{AB}\@+\@
\delta\@^{HK}_{PB}\@N\@^P{}_{HQ}\,N\@^Q{}_{KA}\@+\@T_A\@T_B\Big)\,+                 \\
+\,\h{\nabla}_{\!\DE_A}\Big(N\@^A{}_P{}^P\,-\,N\@^P{}_P{}^A\,\Big)\hskip.8cm
\end{multline}
The last term in eq.~(\ref{3.11}) is the divergence of a vector field $\@X\@$ on $\@M$, with
components $\@X^A:=N\@^P{}_P{}^A\,-\, N\@^A{}_P{}^P\@$\vspace{1pt} independent of the
variable $\@u\@$.

\noindent
In coordinates, recalling eqs.~(\ref{3.8}\@a, b) this implies the exactness relation
\begin{equation*}
\h{\nabla}_{\!\@\DE_A}\@X^A\@\@\sqrt{|g|}\;d\/x^1\vv d\/x^4\,=\,\de/de{x^i}\left(
\sqrt{|g|}\,X^i\right)d\/x^1\vv d\/x^4\,=\,d\left(X\interior\pi^*\/(\eps)\right)
\end{equation*}

\smallskip\noindent
Once again, up to unessential contributions, we are thus left with the expression
\begin{equation}\label{3.12}
I\/(\?\Xi\?)\@=\int_{\psi\/(D)}\@g^{AB}\/\Big(\h{R}_{AB}\@+\@\delta\@^{HK}_{PB}
\@N\@^P{}_{HQ}\,N\@^Q{}_{KA}\@+\@T_A\@T_B\Big)\sqrt{|g|}\;d\/x^1\vv d\/x^4\;\;
\end{equation}
Due to this fact, the analysis of the action principle $\@\delta I=0\@$ may be carried on
along the same lines illustrated in \S\;2\@. Partly from this and partly by inspection of
eqs.~(\ref{3.10}), (\ref{3.12}), we derive the following conclusions:
\begin{itemize}
\item
the value of the functional $\?I\@$ is invariant under arbitrary deformations of the section
$\@\psi\@$. Therefore, the requirement $\@\delta\@I\@=\@0\@$ does not pose any condition on
the choice of $\@\psi\@$, consistently with the interpretation of the latter as a
\emph{gauge field\/};
\item
the variation of the right hand side of eq.~(\ref{3.12}) under arbitrary deformations of the
components $\@N^A{}_{BC}\@$ takes the form (\ref{2.18}), with all indices written in
uppercase. The requirement $\@\delta\@I\@=\@0\@$ is therefore equivalent to the condition
$\@N^A{}_{BC}=0\@$, i.e.~to the identification $\@\nabla=\h\nabla\/$;
\item
in order to express the deformation of the metric in the non--holonomic basis
$\{\@\DE_A\@,\@\w^A\@\}\@$ care must be taken of the fact that the basis itself gets
modified by the deformation. To account for this fact, we start with the representation
(\ref{3.6}\@b)\@. From the latter we get the relation
\begin{equation*}
\delta\left(g^{AB}\@\DE_A\Per \DE_B\right)=\delta\?g^{ij}\,\DE_i\Per \DE_j\,+\,
g^{ij}\left(\delta\@\DE_i\Per
\DE_j\@+\@\DE_i\Per\delta\@\DE_j\right)
\end{equation*}
whence, setting $\@\delta\big(g^{AB}\@\DE_A\Per\DE_B\big):=\delta\@\h\phi^{\,AB}\,\DE_A\Per
\DE_B\@$ and recalling eq.~(\ref{3.5})
\begin{equation}\label{3.13}
\delta\@\h\phi^{\,AB}=\Big<\@\delta\?g^{ij}\,\DE_i\Per \DE_j-
g^{ij}\big(\@\delta\@\g_i\;\DE_u\Per\DE_j\@+\@
\delta\@\g_j\;\DE_i\Per\DE_u\@\big)\@,\,\w^A\Per\w^B\@\Big>\;
\end{equation}
\end{itemize}

\noindent
Comparison of eq.~(\ref{3.13}) with eqs.~(\ref{2.13}), (\ref{3.6}) provides the further
identification
\begin{equation*}
\delta\@\sqrt{|g|}\,=\,-\@\frac12\,g_{ij}\;\delta\?g^{ij}\,=\,
-\@\frac12\,g_{AB}\;\delta\@\h\phi^{\,AB}
\end{equation*}
The variation of the functional $\/I\/$ under admissible deformations of the metric takes
therefore the form
\begin{equation*}
\delta\@I=\int_{\psi\/(D)}\/\left[R_{AB}+T_A\@T_B-\frac12\@\left(R+T_P\@T^P\@\right)g_{AB}
\right]\/\delta\@\h\Phi^{\@AB}\,\sqrt{|g|}\;d\/x^1\vv d\/x^4
\end{equation*}
with $\@\delta\@\h\phi^{\,AB}\@$ given by eq.~(\ref{3.13}). Collecting all results we
conclude
\begin{Proposition}\label{Pro3.1}
A necessary condition for a section $\@\Xi:\V4\to\h\R\/(M)\times_M\@\C\/(M)\@$ to be an extremal
for the functional (\ref{3.10}) under arbitrary deformations of all fields
$\@\psi,\h\Phi,\nabla\@$ is the validity of the relations
\begin{subequations}\label{3.14}
\begin{align}
 & \nabla=\h\nabla                                                      \\[2pt]
 & \h R_{0j}\,=\,\h R_{j0}\,=\,0                                        \\
 & \h R_{ij}\,-\,\frac12\left(\h R\@^0{}_0\@+\@\h R\@^k{}_k\right)g_{ij}\,=\,0
\end{align}
\end{subequations}
\end{Proposition}
As a final step we now rephrase eqs.~(\ref{3.14}\@b, c) in terms of the physical fields,
namely the metric of $\@\V4\@$ and the electromagnetic tensor
$\@F:=d\@\big(\kappa^{-1}\,\psi^*\/(\w\@^0)\@\big)\@=$
$=\@\kappa^{-1}\@\W\@_{ij}\@\w\@^i\v\w\@^j\@$. To this end, we evaluate the curvature
$2$--forms of $\@\hat\nabla\@$ in terms of $\@\W\,^i{}_j\,$ and of the curvature $2$--forms
$\@\rho\@^{*\@i}{}_j\@$ of the riemannian connection over $\@\V4\@$.
\linebreak
On account of eqs.~(\ref{3.5}), (\ref{3.9}\@a, b), a straightforward calculation yields the
result
\begin{alignat}{2}
&\hat\rho\,^i{}_j\,&&=\,d\@\hat\w\@^i{}_j\@+\@\hat\w\@^i{}_r\v\hat\w\@^r{}_j\@+
\@\hat\w\@^i{}_0\v\hat\w\@^0{}_j\@= \nonumber                                   \\[1pt]
&&& =\,\rho^{\?*\,i}{}_j\@-\@\big(\@\W\@^i{}_j\,\W\@_{rs}\@+\@\W\@^i{}_r\,\W\@_{js}\@\big)\,
\w\@^r\v\w\@^s\@-\@\big(\@d\@\W\@^i{}_j\@+\@\w\?^{*\@i}{}_p\,\W\@^p{}_j\@-\@
\w\?^{*\@p}{}_j\,\W\@^i{}_p\@\big)\v\w^{\@0}   \nonumber                        \\[5pt]
&\hat\rho\,^0{}_j&&=\,d\@\hat\w\@^0{}_j\@+\@\hat\w\@^0{}r\v\hat\w\@^r{}_j\,=\,
\big(\@d\@\W\@_{jr}\@-\@\w\?^{*\@k}{}_j\,\W\@_{kr}\@\big)\v\w^{\@r}\,-\,
\W\@^k{}_j\@\W\@_{kr}\,\w^{\@r}\v\w^{\@0}       \nonumber                       \\[5pt]
&\hat\rho\,^j{}_0&&=\,d\@\hat\w\@^j{}_0\@+\@\hat\w\@^j{}r\v\hat\w\@^r{}_0\,=\@-\,
\big(\@d\@\W\@^j{}_r\@-\@\w\?^{*\@j}{}_k\,\W\@^k{}_r\@\big)\v\w^{\@r}\,-\,
\W\@^j{}_k\@\W\@^k{}_r\,\w^{\@r}\v\w^{\@0}       \nonumber                       \\[5pt]
&\hat\rho\,^0{}_0&&=\,\@\hat\w\@^0{}r\v\hat\w\@^r{}_0\,=\@-\,
\W\@_{pr}\@\W\@^p{}_s\,\w^{\@r}\v\w^{\@s}          \nonumber
\end{alignat}

\noindent
From this, resuming the standard notation of tensor calculus on $\@\V4\@$ ($\@R_{ij}\@$ for
the Ricci tensor, $\,w^i{}_{j\cdots h\@\|\@k}\;$ for the covariant derivative, etc.), and
recalling eq.~(\ref{2.8}) we get the identifications
\begin{equation}\label{3.15}
\hat R\@_{ij}\,=\,R\@_{ij}\@-\@2\,\W\@^p{}_i\,\W\@_{pj}\,,\qquad \hat R\@_{i\?0}\,=\,
\hat R\@_{0\@i}\,=\,\W\@_i\,^p{}_{\|\,p}\,,\qquad \hat R\@_{0\?0}\,=\,\W\@^{rs}\,\W\@_{rs}
\hskip.4cm
\end{equation}
Collecting all results, and writing $\@\kappa\@F_{ij}\@$ in place of $\@\W_{ij}\@$ we
conclude that, with the ansatz $\@\kappa=\frac{\sqrt{4\?\pi\@G}}{c^2}\@$, eqs.~(\ref{3.14})
are identical to the Einstein Maxwell equations
\begin{equation*}
\left\{
\begin{aligned}
 & F\@_i\,^p{}_{\|\,p}\,=\,0                                             \\[4pt]
 & R_{ij}\@-\@\@\frac12\,R\,g_{ij}\,=\,\frac{8\?\pi\@G}{c^4}\,
 \Big(\/F\@^p{}_i\,F_{pj}\@-\@\frac14\;F\@^{rs}\@F_{rs}\,g_{ij}\@\Big)
\end{aligned}
\right.
\end{equation*}

\smallskip
Once again, it is worth remarking that all previous conclusions hold unchanged if part of
the relations expressed by the Euler--Lagrange equations (\ref{3.14}\@a, b, c) are imposed
as a priori constraints. In particular, if the requirement $\@\nabla=\h\nabla\@$ is assumed
at the outset --- thus giving up the affine degrees of freedom and regarding the dynamical
fields as sections $\@\Upsilon:\V4\to\R\/(M)\@$ in the sense illustrated at the beginning of
this Section --- the functional (\ref{3.10}) reduces to
\begin{equation*}
I\@=\int_{\psi\/(D)}\@g^{AB}\,\h{R}_{AB}\@\sqrt{|g|}\;d\/x^1\vv d\/x^4\;\;
\end{equation*}
Recalling eqs.~(\ref{3.6}\@b), (\ref{3.15}), and evaluating everything in terms of the
physical fields $\@\Phi\@$ and $\@F=\kappa^{-1}\@\W\@$, the latter expression may be written
in the form
\begin{multline}\label{3.16}
I\,=\,\int_D\@\psi^*\left(\@g^{AB}\,\h{R}_{AB}\right)\@\sqrt{|g|}\;d\/x^1\vv d\/x^4\,= \\
=\,\int_D\@\left[\@R\@-\,\frac{4\?\pi\?G}{c^4}\,F_{ij}\,F^{ij}\@\right]
\sqrt{|g|}\;d\/x^1\vv d\/x^4\hskip.4cm
\end{multline}
Under the stated circumstance, the requirement $\@\delta\@I=0\@$ is therefore identical to
the action principle for the Einstein--Maxwell equations in General Relativity.

\appendix
\section{Lie derivative of connections}
Let $M\/$ be an $n$--dimensional differentiable manifold. We denote by
$L\/(M)\xrightarrow{\pi}M\/$ the frame bundle of $M$, and by $\@r_\a\/(\z)=\z \cdot\a\@$ the
right action of $GL\/(n,\Re)$ on $L\/(M)\@$.

\noindent
Given any local chart $\@(U,\coord xn)\@$ in $M\/$, we refer $\@L\/(M)\@$ to fiber
coordinates $\@x^i,y^i{}_j\@$ according to the prescriptions
\begin{equation*}
x^i\/(\z)\,=\,x^i\/(\pi(\z)\?)\,,\quad \z\@_i\,=
\,\Big(\de/de{x^j}\Big)_{\pi(\z)}\,y^j{}_i\/(\z)\qquad\;
\forall\;\z=\left\{\z_1\@,\ldots,\@\z\@_n\right\}\in\pi^{-1}\/(U)
\end{equation*}
In these coordinates, the Lie algebra associated with the action of $\@GL\/(n,\Re)\@$ on
$\@L\/(M)\@$ is spanned by the vector fields
\begin{equation}\label{A.1}
\mathfrak X\@_p{}\@^j\,=\,y\@^i{}_p\;\de/de{\@y\@^i{}_j}
\end{equation}
commonly referred to as the \emph{fundamental vector fields\/} of $\@L\/(M)\@$.

\smallskip
A vector valued $1$--form $\@\l\@^i{}_{j\,\cdots\,k}\@=\@\l\@^i{}_{j\,\cdots\,k\@r}\,
\@d\?x^r\@+\@\l\@^i{}_{j\,\cdots\,k\@a}\@^b\,\@d\?y\@^a{}_b\@$ over $\@L\/(M)\@$ is called
\emph{pseudo--tensorial\/} if and only if it obeys the transport law \cite{KN}
\begin{equation}\label{A.2}
r_\a^{\;\;*}\big(\l\@^i{}_{j\,\cdots\,k}\@\big)\,=\,
\left(\a^{-1}\right)^i{}_p\;\a^q{}_j\,\cdots\,\a^r{}_k\;\,\l\@^p{}_{q\,\cdots\,r}
\qquad\quad\forall\;\a\in GL\/(n,\Re)
\end{equation}
The definition is immediately extended to vector valued $r$--forms. A semibasic
pseudo--tensorial $r$--form is called \emph{tensorial\/}. The reason for this denomination
is that, given any tensorial $r$--form $\@\l\@^i{}_{j\,\cdots\,k}\@$, the vector valued
function
\begin{equation*}
\l\@^i{}_{j\@\cdots\@k\@\?b_1\@\cdots b_r}\,:=\,
y^{a_1}{}_{b_1}\@\cdots\@y^{a_r}{}_{b_r}
\left<\@\l\@^i{}_{j\@\cdots\@k}\;\bigg|\;\de/de{x^{a_1}}\vv\de/de{x^{a_r}}\@\right>
\end{equation*}
defines a tensor field over $\@M$, whose components in any basis $\@\z\@$ coincide with the
values $\@\l\@^i{}_{j\@\cdots\@k\@\?b_1\@\cdots b_r}\/(\z)\@$.

Every vector field $\@X\in\D^1\/(M)\@$ may be lifted to a field $\@\tilde
X\in\D^1\/(L\/(M))\@$, related in an obvious way to the push forward of the $1$--parameter
group of diffeomorphisms induced by $X\/$. The operation, described in coordinates as
\begin{equation}\label{A.3}
X=X^i\@\de/de{x^i}\quad\longrightarrow\quad\tilde X\,=\,X^i\,\de/de{\@x^i}\,+\,\de
X^i/de{\@x^k}\;y\@^k{}_j\;\de/de{\@y\@^i{}_j}
\end{equation}
is called the \emph{universal lift\/} of vector fields. By construction, the field $\@\tilde
X\@$ is invariant under the action of $GL\/(n,\Re)$, as confirmed by the commutation
relations
\begin{equation}\label{A.4}
\Big[\@\tilde X\@,\,\mathfrak X\@_p{}\@^j\,\Big]\,=\,\left[\@
X^r\,\de/de{\@x^r}\,+\,\de X^r/de{\@x^k}\;y\@^k{}_s\;\de/de{\@y\@^r{}_s}\;,\;\;
y\@^i{}_p\;\de/de{\@y\@^i{}_j}\,\right]\,=\,0
\end{equation}
Due to this fact, the $1$--parameter group of diffeomorphisms associated with $\tilde X\/$
\emph{commutes\/} with the action of $GL\/(n,\Re)$. Given any pseudo--tensorial $r$--form
$\@\l\@^i{}_{j\,\cdots\,k}\@$, the Lie derivative $\@\L_{\tilde
X}\;\l\@^i{}_{j\,\cdots\,k}\@$ is therefore once again pseudo--tensorial.

By definition, a connection $\@\nabla:M\to \C\/(M)\,$, locally described by connection
$1$--forms $\w\@^k{}_j:=\GG(ij,k)\@d\/x^i\/$ is a \emph{horizontal distribution\/} in
$\@L\/(M)\@$, identified with the annihilator of the vector--valued pseudo--tensorial
$1$--form\@{} \cite{KN}
\begin{equation}\label{A.5}
\tilde\w\@^a{}_b\,=\,(y^{-1})^a{}_r\left(d\?y^r{}_b\@+\@\w\@^r{}_s\,y\@^s{}_b\right)
\end{equation}
In view of our previous observations, given any vector field $\@X\@$ on $\/M$, the Lie
derivative $\@\L_{\tilde X}\;\tilde\w\@^a{}_b\,$ along the universal lift of $\@X\@$ is then
a pseudo--tensorial $1$--form over $\@L\/(M)\@$. Moreover, eqs.~(\ref{A.2})\@--\@(\ref{A.4})
imply the relation
\begin{equation*}
\Big<\@\L_{\tilde X}\;\tilde\w\@^a{}_b\;,\;\mathfrak X\@_p{}\@^j\@\Big>\,=\,
\tilde X\left(\Big<\@\tilde\w\@^a{}_b\,,\,\mathfrak X\@_p{}\@^j\@\Big>\right)-\,
\Big<\@\tilde\w\@^a{}_b\,,\,\left[\@\tilde X\@,\,\mathfrak X\@_p{}\@^j\,\right]\Big>\,=\,
\tilde X\Big(\delta\@^a_p\,\delta\@^j_p\@\Big)\,=\,0
\end{equation*}
showing that $\@\L_{\tilde X}\;\tilde\w\@^a{}_b\,$ is also semibasic, and has therefore a
\emph{tensorial\/} character. As such, $\@\L_{\tilde X}\;\tilde\w\@^a{}_b\,$ defines a
tensor field of type $(1,2)\@$ over $\/M$, henceforth denoted by $\@\L_X\,\nabla\@$, and
called the \emph{Lie derivative\/} of the connection $\@\nabla\@$ along $\@X\/$.

In particular, if the local coordinates are chosen consistently with the requirement
$\@X=\de/de{\@\plus{7.5}0x^1}\@$, eqs.~(\ref{A.3}), (\ref{A.5}) provide the relation
\begin{equation*}
\L_{\tilde X}\;\tilde\w\@^a{}_b\,=\,\L_{\de/de{\plus60x^1}}\;\tilde\w\@^a{}_b\,=\,
\big(\@y^{-1}\big)^a{}_r\;y\@^s{}_b\;\de\@\GG(ks,r)/de{\@\plus{7.5}0x^1}\;d\?x^k
\end{equation*}
mathematically equivalent to the representation
\begin{equation}\label{A.6}
\L_X\,\nabla\,=\,\de\@\GG(ks,r)/de{\@\plus{7.5}0x^1}\;d\?x^k\Per d\?x^s\Per\de/de{x^r}
\end{equation}
Therefore, under the stated circumstance, $\@\L_X\,\nabla=0\@$ if and only if
$\;\de\@\plus03\GG(ks,r)/de{\@\plus{7.5}0x^1}=0\@$.

More generally, in arbitrary coordinates, denoting by $\@T\@^i{}_{jk}\@,\,R\@^i{}_{jkl}\@$
and $\@\|\@$ respectively the torsion tensor, the curvature tensor and the covariant
derivative associated with $\@\nabla\@$, a straightforward but lengthy calculation yields
the result
\begin{equation*}
\L_{\tilde X}\;\tilde\w\@^a{}_b\,=\,\big(\@y^{-1}\big)^a{}_r\;y\@^s{}_b\left[\@\left(\@
X\@^r{}_{\|\@s}\@+\@X\@^p\,T\@\@^r{}_{ps}\@\right)_{\|\@k}\,+
\,X\@^p\@R\@\@^r{}_{spk}\@\right]d\?x^k
\end{equation*}
corresponding to the representation
\begin{equation}\label{A.7}
\L_X\,\nabla\,=\,\left[\@\left(\@
X\@^r{}_{\|\@s}\@+\@X\@^p\,T\@\@^r{}_{ps}\@\right)_{\|\@k}\,+
\,X\@^p\@R\@\@^r{}_{spk}\@\right]d\?x^k\Per d\?x^s\Per\de/de{x^r}
\end{equation}
As an indirect check, the reader may verify that eq.~(\ref{A.7}) reduces to eq.~(\ref{A.6})
whenever the condition $\@X=\de/de{\@\plus{7.5}2x^1}\@$ is satisfied. The validity of
eq.~(\ref{A.7}) in any coordinate system is then ensured by the tensor character of both
sides.

\end{document}